\documentstyle[12pt,epsfig,cite,amsbsy]{article}
\def\thebibliography#1{\list    
 {[\arabic{enumi}]}{\settowidth\labelwidth{[#1]}\leftmargin\labelwidth
 \advance\leftmargin\labelsep
 \usecounter{enumi}}
 \def\newblock{\hskip .11em plus .33em minus -.07em}
 \sloppy
 \sfcode`\.=1000\relax}

\begin{document}

\bibliographystyle{unsrt}

\begin{titlepage}

\vspace*{-0.5cm}

\LARGE  
\begin{flushright}
\Large 
HUB-EP-99/47\\
{\large August 25, 1999}
\end{flushright}
\normalsize                          

\vspace*{-1cm}

\vspace*{3cm}

\vspace*{0.5cm}

\LARGE
\begin{center}        
{\bf                      
\boldmath
Can the sneutrino be the lightest supersymmetric particle ?
}
\end{center}                          

\vspace*{1.15cm} 

\Large         
\begin{center}  
Thomas Hebbeker
\end{center}   

\large          
\begin{center} 
Institut f\"ur Physik der               
Humboldt-Universit\"at zu Berlin,
\\
Invalidenstr. 110, D-10115 Berlin, Germany
\\
www.physik.hu-berlin.de
\end{center}

\vspace*{0.5cm}

\vspace*{4.5cm}

\Large
\begin{center}
Abstract
\end{center}

\large  
Within the framework of the 
constrained 
Minimal Supersymmetric extension of the 
Standard Model
we show that recent LEP I limits on the invisible Z width  
exclude the possibility 
that the lightest sparticle is the sneutrino.

\end{titlepage}

\normalsize
\newpage   

\setcounter{page}{2}

\underline{\bf Introduction}

We investigate the constrained 
`Minimal Supersymmetric extension of the 
Standard Model' (MSSM)
as used for example by the LEP collaborations\cite{pdg}.
It assumes Grand Unification, no extra CP violation,
a common scalar mass scale, etc., so that out of more than 100 possible
new constants in a general SUSY model 
only the following free parameters
are left:
\begin{itemize}
\item $m_0$ = Universal scalar mass at the GUT scale 

\item $M_2$ = $SU(2)$ gaugino mass at the electroweak scale 

\item $\mu$ = Higgs(ino) mass parameter (elw. scale)

\item $\tan \beta$ = Ratio of higgs
vacuum expectation values (elw. scale)

\end{itemize}
The additional parameters $A_0$ and 
$m_A$ are not important here.
The quantum number 
R Parity 
is assumed to be conserved, so that the lightest
supersymmetric particle (LSP) is stable.
Cosmological arguments together with limits on abundances of
atoms with anomalous charge over mass ratios 
require that the 
LSP carries neither 
colour nor electrical charge\cite{ellis1}.

In the 
MSSM 
only 
two particles fulfil these constraints: The lightest neutralino,
$\tilde \chi_1^0$, and the sneutrino $\tilde \nu$.
Note that the common scalar mass $m_0$ implies that the three sneutrinos
$\tilde \nu_e, \tilde \nu_{\mu}, \tilde \nu_{\tau}$
are degenerate in mass, and we 
do not 
distinguish between them.  
A third LSP candidate is the gravitino, but in the constrained
MSSM it is  assumed 
to be heavier than the other SUSY particles, as predicted
in supergravity models.

A priori it is not clear which one is the LSP. Since the existing 
upper mass limit for the sneutrino is better than for the lightest
neutralino 
\cite{hebbeker}, many physicists concentrate
on the hypothesis LSP = $\tilde \chi_1^0$.
In this paper we investigate for which SUSY parameters 
the sneutrino plays the role of the LSP, and to what extent 
this possibility is ruled out by existing experimental bounds.

\bigskip

\underline{\bf Limit on sneutrino mass}

First we analyse the experimental bounds; it turns out that the 
limit obtained in $e^+e^-$ collision experiments
with centre of mass energies around the Z pole
is most stringent \cite{pdg}.

The LEP I measurements of the Z properties 
allow to constrain the non Standard Model 
contributions to the invisible Z width to
\cite{tampere}
\begin{eqnarray}
\Delta \Gamma_{\mathrm{inv}} < 2.0 \, \mathrm{MeV}  \;\;\;\;\;\;
95 \% \, CL \;\;  , 
\label{eq0}
\end{eqnarray} 
assuming 3 light neutrino species. 
`Invisible' decay channels are those, for which a substantial
fraction (typ. 50\% or more) 
of the energy carried by the final state particles 
is unseen in the detector and which 
are inconsistent with fermion pair production.
Also sneutrino pairs might be produced in Z decays. 
If they act as LSP they are stable and undetected, thus
contributing to $\Gamma_{\mathrm{inv}}$. 
For the conclusions of this paper it is sufficient to discuss this
case.

The sneutrino contribution to the invisible Z width is given by\cite{haber}:
\begin{eqnarray}
\Delta \Gamma_{\mathrm{inv}}^{\tilde \nu}  = 3 \cdot \frac{1}{2} 
\cdot \left[ 1- \left( \frac{2 m_{\tilde \nu}}{m_Z} \right)^2 \right] ^{3/2} 
\cdot \Gamma_{\mathrm{inv}}^{\nu}
\end{eqnarray}
Here $\Gamma_{\mathrm{inv}}^{\nu} = 167 \, \mathrm{MeV}$ 
is the neutrino contribution for one
family.
The factor 3 stands for the 3 families, $\frac{1}{2}$ 
results from the different spins of neutrinos and sneutrinos, and
the term in brackets containing the sneutrino mass describes the 
kinematical suppression.

The experimental upper limit (\ref{eq0})
can be converted into a sneutrino mass limit of
\begin{eqnarray}
m_{\tilde \nu}^{LSP} > 44.6 \, \mathrm{GeV}  \;\;\;\;\;\;
95 \% \, CL
\end{eqnarray}

%
%
%
%
%
%
%
%
%
%
%

%
%
%
%
%
%
%
This bound 
improves the older limit of $43.1 \, \mathrm{GeV}$\cite{ellis2,pdg}.

\medskip

It should be noted that
our limit holds 
also
in the 
more general case 
that either 
$\tilde \nu$ {\bf or} $\tilde \chi_1^0$ act as the LSP.
In the latter case the sneutrino will decay.
If it is long lived, it escapes
detection.
If it is short lived the two dominant 
decay modes are neutrino plus neutralino and
lepton plus chargino\cite{haber}.
In the first case
all or a large 
fraction of the energy escapes undetected.
The second case is already ruled out from 
the lower limit on the
chargino mass of $m_Z/2$, derived 
from the {\bf total} Z width measured at LEP I\cite{pdg}.

\bigskip

\underline{\bf Sneutrino-LSP in the MSSM}

Now we turn to the sparticle masses as predicted in the constrained
MSSM and investigate if we can set a theoretical upper limit 
on $m_{\tilde \nu}$.

To be the LSP the sneutrino mass
must in particular fulfil the two
relations
\begin{eqnarray}
m_{\tilde \nu} & < & m_{\tilde e_R} 
\\
m_{\tilde \nu} & < & m_{\tilde \chi_1^0} 
\end{eqnarray}
which are
{\bf not} true in large regions of the MSSM parameter space.
Note that $m_{\tilde e_L} > m_{\tilde e_R}$ is always fulfilled.
The charged sleptons $\tilde \mu$ and $\tilde \tau$
are heavier than $\tilde e_R$
(with the stau possibly making an exception, if mixing is large;
this would lead to the additional constraint 
$ m_{\tilde \nu} < m_{\tilde \tau} $,
yielding an even
better sneutrino mass limit than the one presented below).  

\medskip

To understand the 
first relation (4) we calculate the two slepton masses 
using the approximate formulae given in \cite{martin}:
\begin{eqnarray}
m_{\tilde \nu}^2  
& = & m_0^2 - 0.5 \, m_Z^2 \, \frac{\tan^2 \beta-1}{\tan^2 \beta + 1}   
+ 0.80 \, M_2^2
\\  
m_{\tilde e_R}^2  - m_e^2 
& = & m_0^2 + \sin^2\theta_W \, m_Z^2 \, \frac{\tan^2 \beta - 1}{\tan^2 \beta + 1}   
+ 0.22 \, M_2^2
\end{eqnarray}
The second term on the right hand side 
is due to quartic sfermion-higgs
couplings. The term proportional to $M_2^2$ 
describes the running of the masses
from the GUT scale to the electroweak scale.
\\
Thus (4) is fulfilled if
\begin{eqnarray}
\frac{\tan^2 \beta -1}{\tan^2 \beta +1} \; 
 > \; 0.79 \, \frac{M_2^2}{m_Z^2}
\label{eq1}
\end{eqnarray}
using $\sin^2\theta_W = 0.23$ and neglecting the electron mass.
Since the left hand side is smaller than 1, we find 
in particular
\begin{eqnarray}
M_2  < 1.13 \, m_Z = 103 \, \mathrm{GeV}
\end{eqnarray}
%
%
Using the 
program SUSYGEN\cite{susygen}, in which
the sparticle masses are calculated more precisely\cite{ambrosanio},
we find a similar bound of $ 104 \, \mathrm{GeV}$.

\medskip

The condition (5) 
is more difficult to understand, since two more 
MSSM parameters come into play:
$m_0$, which determines the sneutrino mass, and 
the higgsino mass parameter $\mu$, 
appearing in the neutralino mass matrix.
Using the basis for the interaction eigenstates
as  given in reference \cite{susygen}, the mass matrix becomes 
\begin{eqnarray}
\left(
\begin{array}{cccc}
0.61 \, M_2   &  0.21 \, M_2  & 0 & 0 \\
0.21 \, M_2 & 0.88 \, M_2 & m_Z & 0 \\
0 & m_Z & \mu \, \sin 2 \beta & - \mu \, \cos 2 \beta \\
0 & 0  & - \mu \, \cos 2 \beta & - \mu \, \sin 2 \beta \\
\end{array}
\right)
\end{eqnarray}
Here the GUT gaugino mass relations and the 
numerical value for the weak mixing angle have been used.

The smallest eigenvalue, the neutralino mass $m_{\tilde \chi^0_1}$,
can become large only if both $M_2$ {\bf  and}  $|\mu|$ are large. 
Equation (9) therefore implies an upper bound on 
$m_{\tilde \chi^0_1}$ and, through (5), on $m_{\tilde \nu}$, of the order
of $m_Z$. 

After these qualitative arguments we need to 
determine the upper limit on the 
LSP sneutrino mass quantitatively.
We computed $m_{\tilde \nu}$ for many points
in the MSSM parameter space and calculated
the maximum mass value from the subset of 
points which respect (4) and (5). 

First we used the mass
formulae as given above and 
diagonalised the neutralino mass matrix 
numerically.
The parameter space was 
scanned in the range $0 < M_2 < 110 \, \mathrm{GeV}$,
$0 < m_0 < 1000 \, \mathrm{GeV}$,
$ \pm \mu  < 1000 \, \mathrm{GeV}$,
$ 1 < \tan \beta < 50$.
The characteristic value of
$1000 \, \mathrm{GeV}$ is motivated by the 
requirement that SUSY solves the hierarchy problem.  
More than 1 billion points have been considered.
Result: $ m_{\tilde \nu}^{LSP} < 44.3 \, \mathrm{GeV}$.

We repeated the procedure 
with SUSYGEN, which is more precise but less fast.
In order to save computer time,
we scanned only through that subset of the MSSM parameters
for which 
the approximate formulae predict
high values of 
$ m_{\tilde \nu}^{LSP}$.
The step sizes were $0.1\, \mathrm{GeV}$ in $M_2$ and $m_0$,
$0.1$ in $\tan \beta$ and $5 \, \mathrm{GeV} $ in $\mu$ 
(on which the sneutrino mass depends only indirectly).
The resulting theoretical upper limit is
\begin{eqnarray}
 m_{\tilde \nu}^{LSP} < 44.2 \, \mathrm{GeV}
\end{eqnarray}
in good agreement with the approximate value of $44.3 \, \mathrm{GeV}$.
The corresponding MSSM parameters  are 
$M_2 = 84.1 \, \mathrm{GeV}$, 
$m_0 \to 0 $, 
$\tan \beta =4.2$
and $\mu \approx - 190 \, \mathrm{GeV} $. 
%
%
%
%
The neutralino mass is nearly degenerate with the sneutrino mass in this
case.

The difference between the
experimental and theoretical limits on the sneutrino mass derived
in this paper is rather small. Therefore
the inclusion of higher oder corrections both to 
the sneutrino contribution to the Z width as well as to the 
sparticle masses is desirable.

An improved experimental limit cannot be expected in the near future.
The LEP I data taking and analyses are completed, and at LEP II
the cross section for the relevant channel, 
$\mathrm{e^+e^-} \, \to \, \tilde \nu \, \bar {\tilde \nu} \; \gamma 
$, 
is 
small. 


\bigskip

\underline{\bf Conclusions}

LEP I data show that 
the sneutrino must be heavier than $44.6 \, \mathrm{GeV} $
at the $95\% $ confidence level.
In the sneutrino LSP scenario
this experimental lower bound 
is inconsistent with 
the theoretical upper limit on the sneutrino mass. 
Therefore - within the constrained MSSM -
the sneutrino can {\bf not} be the LSP!

\newpage

 
\underline{\bf Acknowledgements}
 
We would like to thank Martin Gr\"unewald,  
Christian Preitschopf and 
Daniel Ruschmeier
for valuable comments. 

\bigskip


\underline{\bf References}

\bibliography{sneutrino}

\end{document}